\documentclass[twocolumn]{el-author}

\newcommand{\ua}{\uparrow}
\newcommand{\nc}{\newcommand}
\nc{\da}{\downarrow} \nc{\hc}{\hat{c}} \nc{\hS}{\hat{S}}
\nc{\bra}{\langle} \nc{\ket}{\rangle} \nc{\eq}{equation (\ref}
\nc{\h}{\hat} \nc{\hT}{\h{T}}\nc{\be}{\begin{eqnarray}}
\nc{\ee}{\end{eqnarray}}\nc{\rd}{\textrm{d}}\nc{\e}{eqnarray}\nc{\hR}{\hat{R}}\nc{\Tr}{\mathrm{Tr}}
\nc{\tS}{\tilde{S}}\nc{\tr}{\mathrm{tr}}\nc{\8}{\infty}\nc{\lgs}{\bra\ua,\phi|}\nc{\rgs}{|\ua,\phi\ket}
\nc{\hU}{\hat{U}}\nc{\lfs}{\bra\phi|}\nc{\rfs}{|\phi\ket}\nc{\hZ}{\hat{Z}}\nc{\hd}{\hat{d}}\nc{\mD}{\mathcal{D}}
\nc{\bd}{\bar{d}}\nc{\bc}{\bar{c}}\nc{\mc}{\mathcal}\nc{\ea}{eqnarray}\nc{\mG}{\mathcal{G}}\nc{\bce}{\begin{center}}
\nc{\ece}{\end{center}}
\usepackage{fancyhdr}
\fancypagestyle{plain}{%
   \fancyhf{} 
   \fancyfoot[C]{\scriptsize This paper is a postprint of a paper submitted to and accepted for publication in ELECTRONICS LETTERS and is subject to Institution of Engineering and Technology Copyright. The copy of record is available at IET Digital Library. DOI: 10.1049/el.2016.2864.}
   
}

\begin{document}

\title{Optimization of Chaos-based Coded Modulations for Compensation of Amplifier Nonlinearities}

\author{Francisco J. Escribano, Jos\'e S\'aez--Landete and Alexandre Wagemakers}

\abstract{In this work we expand the possibilities of a known class of chaos-based coded modulation (CCM) systems to address the problem of high-power amplifier (HPA) nonlinearity. The hypothesis is that the CCM nonlinear nature can be exploited to counteract its effects on the system performance, avoiding the need of a predistorter. We propose an optimization method for the design of the CCM to prove this hypothesis. The results show that, for a given back-off level, a nonlinear mapping function for the chaos-based encoder and decoder can be designed to compensate the HPA effect on the error rate.}

\maketitle

\section{Introduction}

Chaos-based communication methods have been witnessing a growing interest. Although they may not be clearly advantageous in strictly linear channels, their inherent nonlinear nature may make them well fitted to nonlinear channels. One great concern in mobile devices design has to do with the need to increase battery life, and this normally requires optimizing the power usage of the RF stage. This leads to set the RF amplifier near its saturation point, where the response is highly nonlinear. In single carrier scenarios, the related undesirable effects are usually compensated by predistortion.

In the present work, we take advantage of the structure of a family of chaos-based coded modulated (CCM) communication schemes that make use of their symbolic dynamics to encode and decode the information. These systems are based on chaotic piecewise linear maps (PWLM) and are equivalent to finite-state machines. They may change their output samples probability density function (pdf) by using a nonlinear conjugation function \cite{Escribano06a}. The hypothesis is that, by appropriately managing and optimizing this conjugation function, we may counteract the undesirable effects of the RF high-power amplifier (HPA) on the bit error probability (BEP), avoiding the need of a predistorter.

\section{System setup}

The chaotic encoder is based on the already proposed model for PWLM CCM blocks \cite{Escribano10a}. The basic CCM encoding block produces chaos-coded samples $z_n\in\left[0,1\right]$
\begin{equation}
 \label{eq:encoding}
 z_n=f\left(z_{n-1},b_n\right)+b_n\cdot 2^{-Q},
\end{equation}
where $f\left(\cdot,\cdot\right)$ is the chaotic (multimap) PWLM function, the second part is the small perturbation allowing for the introduction of the information bit $b_n$ in the symbolic dynamics of the system for a quantization factor of $Q$ bits, and $n$ is the time index. For details, please see \cite{Escribano10a}.

A conjugation function, denoted as $h\left(z\right):\left[0,1\right]\rightarrow\left[0,1\right]$, is used to change the pdf of the chaotic samples \cite{Escribano06a}. This conjugation function has to be strictly non-decreasing, with $h\left(0\right)=0$, $h\left(1\right)=1$. Its output is then normalized to have zero-mean, as
\begin{equation}
 \label{eq:conjugation}
 x_n=2\cdot h\left(z_n\right)-1=2\cdot s_n -1.
\end{equation}

The chaos encoded sequence $x_n$ is fed to an HPA, based on the Saleh model \cite{Saleh81}, and goes through an additive white Gaussian noise (AWGN) channel. At the input of the decoder, we have
\begin{equation}
 \label{eq:received}
 r_n= y_n + n_n=A \cdot g_{NL}\left(x_n\right)+n_n,
\end{equation}
where $g_{NL}\left(\cdot\right)$ represents the AM/AM distortion function of the Saleh model, $A$ is a scaling factor to keep signal-to-noise ratio unchanged for comparison purposes, and $n_n$ is a sample of the zero-mean AWGN process, with variance $\sigma_n^2$. The BEP will be measured against the $E_b/N_0$ factor, that, under the standard assumption of appropriate pulse shaping and matched filter reception, may be calculated as $E_b/N_0=P/\left(2\sigma_n^2\right)$, where $P=\mathrm{E}\left[x_n^2\right]$ is the power of the chaos encoded sequence. The AM/AM distortion function takes the form
\begin{equation}
 \label{eq:saleh}
 g_{NL}\left(x_n\right)=\frac{\alpha \cdot B \cdot x_n }{1+\beta \cdot |B \cdot x_n |^2},
\end{equation}
where $\alpha=2.1587$, $\beta=1.1517$ are standard values, and $B$ is an input back-off factor controlling the amount of HPA nonlinearity.

The received sequence $r_n$ is then decoded according to a maximum {\it a posteriori} (MAP) algorithm that takes advantage of the finite-state machine equivalence of the system \cite{Escribano06a}. The nonlinear mapping $h\left(\cdot\right)$ will be optimized to compensate the effect of $g_{NL}\left(.\right)$ on the error performance, while keeping the CCM basic structure of \eqref{eq:encoding} unchanged.

\section{Bound calculation}

The objective function of the optimization would ideally be the BEP, but it is not possible to provide a closed-form formula for it. Therefore, it will be approximated by means of a bound, that may be calculated on the basis of the pairwise error probability (PEP) \cite{Escribano10a}
\begin{equation}
 \label{eq:PEP}
 P\left(\mathbf{x}\rightarrow \mathbf{x}'\right | \mathbf{x}, \mathbf{e})= \frac{1}{2} \mathrm{erfc}\left(\sqrt{\frac{d^2_{eq}}{4P}\frac{E_b}{N_0}} \right),
\end{equation}
for a specific encoded sequence represented by vector $\mathbf{x}$, and incorrectly chosen sequence $\mathbf{x}' \neq \mathbf{x}$. Both sequences are linked in the trellis through a binary error loop $\mathbf{e}$ of length $L\left(\mathbf{e}\right)$. The quantity $d^2_{eq}$ is an equivalent Euclidean distance term calculated as
\begin{equation}
 \label{eq:distance}
 d_{eq}=\frac{\sum_{n=m}^{m+L\left(\mathbf{e}\right)-1} |y_n-x'_n|^2 - \sum_{n=m}^{m+L\left(\mathbf{e}\right)-1} |y_n-x_n|^2}{\sqrt{\sum_{n=m}^{m+L\left(\mathbf{e}\right)-1} |x_n-x'_n|^2}},
\end{equation}
for the error loop of length $L\left(\mathbf{e}\right)$ under consideration, and for the actual $\mathbf{x}$ encoded sequence, diverging from $\mathbf{x}'$ at time index $m$. The details about the derivation of this PEP can be seen in \cite{Escribano10a}. Its adaptation to the present context is straightforward, by correctly identifying the incumbent binary error loops in the encoder trellis. This PEP assumes maximum likelihood (ML) decoding, but it provides a good approximation for the bit error rate (BER) under MAP decoding for sufficiently high $E_b/N_0$.

If we consider, for a specific CCM, a set of most probable error loops $B_{\mathbf{e}}$, the BEP can be estimated through the bound
\begin{equation}
 \label{eq:bound}
 P_b \approx \hat{P}_b \overset{\scriptscriptstyle\Delta}= \sum_{\mathbf{e} \in B_{\mathbf{e}}} \sum_{\mathbf{x}} \frac{\omega\left(\mathbf{e}\right)}{2^{Q+L\left(\mathbf{e}\right)}} \cdot P\left(\mathbf{x}\rightarrow \mathbf{x}'\right | \mathbf{x}, \mathbf{e}),
\end{equation}
where the second summation is extended over all the possible encoded sequences $\mathbf{x}$ of length $L\left(\mathbf{e}\right)$, and where $\omega\left(\mathbf{e}_i\right)$ is the associated binary weight. The possible encoded sequences of length $L\left(\mathbf{e}\right)$, taking into account all the starting states for the trellis, amount to $2^{Q+L\left(\mathbf{e}\right)}$. Please note that no simplification can be performed assuming the uniform error property of other related trellis encoded modulation (TCM) schemes, because of the nonlinear character of the system.

For the CCM systems considered, as the channel impairments have no memory, the error loops that will appear for high $E_b/N_0$ are exactly the same as under AWGN. The most probable error loops, and the affected by lowest $d^2_{eq}$ distances, will thus be the shortest ones, so that we limit the calculation of \eqref{eq:bound} to the error loops of length $L\left(\mathbf{e}\right)$ close to $Q$ \cite{Escribano10a}.

\section{Optimization}

In order to design the CCM system, we propose the optimization of the conjugation function $h\left(\cdot\right)$ with the objective of improving the system BEP, including the nonlinear effects of the HPA. We approximate the BEP by means of the bound $\hat{P}_b$ \eqref{eq:bound}, which depends on the conjugation function through \eqref{eq:conjugation}--\eqref{eq:distance}. In order to provide a numerical approximation of $h\left(\cdot\right)$, we sample its domain $[0,1]$ with $M$ equidistant samples. The final conjugation function may be thus obtained by interpolation techniques. The samples are $z^i=i/M, i=0,\ldots,M$, matched to the function samples $s^i=h\left(z^i\right)$. We use this notation to distinguish these variables from the related temporal sequences $z_n$ and $s_n$. Due to the nature of the conjugation function, the discretized values $s^i$ and $z^i$ must meet the constraints described in the system setup. The formulation of the optimization problem is given by 
\begin{equation}
\label{eq:optimization}
 \begin{array}{clr}
 \displaystyle \min_{s^i} & \lbrace \hat{P}_b \rbrace  & \\
 \textrm{s.t.} 	& z^i = i/M, 	\\
		& 0  <  s^i  <  1, 	\\
		& s^i < s^{i+1},	\\
 \end{array}
\end{equation}
where $i=1,\cdots,M-1$. This problem is a classical minimization of a nonlinear objetive function with linear constraints. We use the Interior Point Algorithm implemented in MATLAB Optimization Toolbox, described in \cite{Byrd00} and \cite{Waltz06}. The optimization converges always to the same solution, even when using different starting points. For practical reasons, we use a linear function as seed. Note that the optimization is performed for the system design, so the processing time is not a problem.

\section{Results and discussion}

The bounds and BER results will be compared with an uncompensated standard system of similar characteristics and equal spectral efficiency, a $4$-PAM setup with a Viterbi soft-decoded $R=1/2$ non-systematic convolutional code (CC), with octal polynolmials $133$ and $171$. The optimization, bound calculation and BER simulation of the chaos-based system are performed for $Q=5$, while the mapping of chaos-based samples through $h\left(\cdot\right)$ is performed using linear interpolation. In the bound calculation, the error loops considered are just those with lengths ranging from $Q$ to $2\cdot Q$. In all the cases, trellis termination is enforced, and the data block takes length $10000$ bits.

The number of samples considered in the optimization procedure is $M=101$, but values above $50$ yield the same solution. The optimization is performed for $E_b/N_0=10$dB, which is a suitable working point. For clarity's sake and without loss of generality, we have just chosen two instances of CCM systems: the one based on the Bernoulli shift map (BSM), and the one based on the multimap version of the tent map (mTM) \cite{Escribano10a}. The BSM represents a kind of CCM that provides no coding gain in AWGN, while the mTM one does. The HPA back-off factor considered is the input back-off.

\begin{figure}[h]
\centering{\includegraphics[width=60mm]{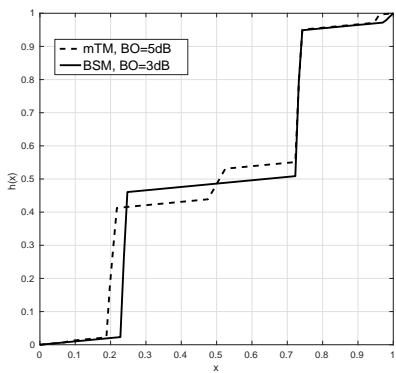}}
\caption{\label{fig:hx}$h\left(\cdot\right)$ for BSM and mTM with HPA in nonlinear regime.}
\end{figure}
We show some results for the optimized $h\left(\cdot\right)$ in Fig. \ref{fig:hx}. We have explored a variety of cases with other back-off factors, and the principles shown remain: $h\left(\cdot\right)$ is made of a number of alternating linear sections with low and high slopes. The reason for this is that such kind of conjugation function produces an output data pdf with high concentration around the output values corresponding to the low slope sections (near $0$, $1/2$ and $1$ for the BSM case shown in Fig. \ref{fig:hx}, for example). Under the presence of the HPA nonlinearity, this creates an improved balance on the spectra of $d_{eq}^2/\left(2P\right)$ values in (\ref{eq:distance}). Note how this contrasts with the results in \cite{Escribano06a}, where, in the linear AWGN channel, no gain was to be got just by forcing the data pdf to have maxima around $0$ and $1$.

\begin{figure}[h]
\centering{\includegraphics[width=81mm]{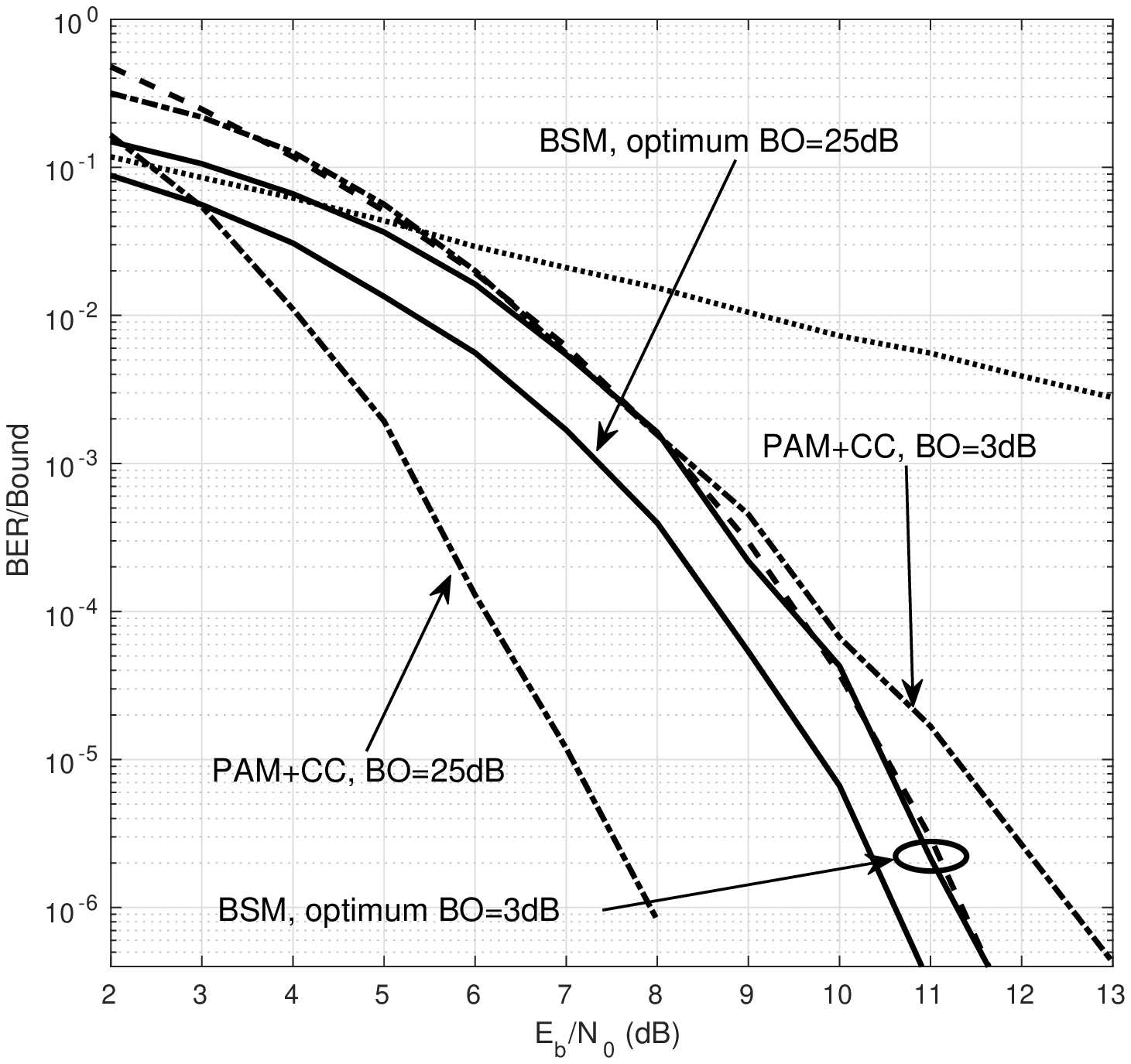}}
\caption{\label{fig:BSM}Solid lines: BER for optimized BSM. Dash-dotted lines: BER for $4$-PAM+CC. Dashed line: bound for optimized BSM with back-off $3$dB. Dotted line: BER of BSM with back-off $3$dB without conjungation function.}
\end{figure}
\begin{figure}[h]
\centering{\includegraphics[width=81mm]{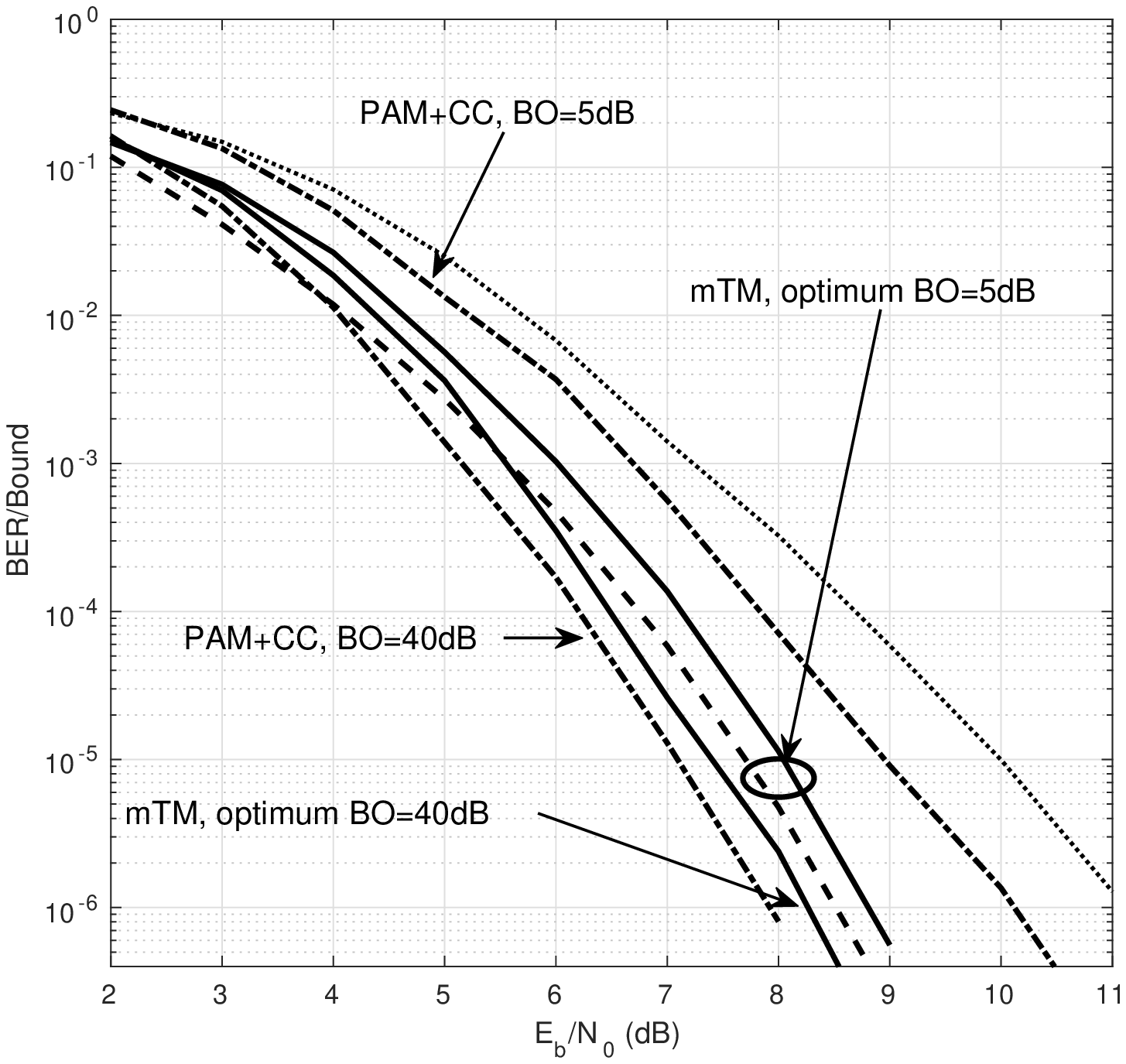}}
\caption{\label{fig:mTM}Solid lines: BER for optimized mTM. Dash-dotted lines: BER for $4$-PAM+CC. Dashed line: bound for optimized mTM with back-off $5$dB. Dotted line: BER of mTM with back-off $5$dB without conjungation function.}
\end{figure}

In Figs. \ref{fig:BSM} and \ref{fig:mTM} we have depicted the results. On the one hand, we may verify a good agreement between the BER and the corresponding bound. In the mTM case, the bound is less tight due to its more irregular nature, as compared to the BSM. Nonetheless, the optimization of the conjugation function based on \eqref{eq:bound} shows to be successful enough: note how, for the highly nonlinear cases ($3$ and $5$dB back-off), there is a steady gain when using the optimized conjugation function with respect to the case of not using it. Moreover, the uncompensated classical counterpart loses around $2$dB $E_b/N_0$ at a BER of $10^{-5}$ when going from $40$dB to $5$dB back-off, or $4$dB when going from $25$dB to $3$dB back-off. In the same situations, the chaos-based systems lose less than $1$dB $E_b/N_0$. This alleviates the need for a predistorter in the CCM case, in contrast to the classical counterpart.

\section{Conclusion}
We have proposed a method to minimize the impact of HPA nonlinearity on the bit error rate of chaos-based coded modulation systems, without the need of a predistorter. The bound used as objective function has shown to be accurate for the cases tested, and the BER obtained with the optimized conjugation function outperforms a similar uncompensated classical alternative, that clearly requires predistorsion. Apart from the RF HPA context, it is to be noted that all this framework can be of interest in any communication system requiring amplitude modulation and containing devices with nonlinear responses, for example, in the new visible light communications systems, where the light emitting diode (LED) has a nonlinear transfer function.

\vskip3pt
\ack{This work has been partially supported by Universidad de Alcala's Research Project CCG2015/EXP-071.}

\vskip5pt

\noindent F. J. Escribano and J. S\'aez (\textit{Universidad de Alcal\'a, Spain}), and A. Wagemakers (\textit{Universidad Rey Juan Carlos, Spain})
\vskip3pt

\noindent E-mail: francisco.escribano@uah.es

\end{document}